\newcounter{myctr}
\def\myitem{\refstepcounter{myctr}\bibfont\noindent\ifnum\themyctr>9\else\phantom{0}\fi\hangindent17pt\themyctr.\enskip}

\documentclass{ws-ijqi}

\begin{document}

\markboth{A. R. Usha Devi, A. K. Rajagopal and Sudha}
{Quantumness of correlations and entanglement}

\catchline{}{}{}{}{}

\title{Quantumness of correlations and entanglement}

\author{A. R. Usha Devi}

\address{Department of Physics, Bangalore University, 
Bangalore-560 056, India.\\
Inspire Institute Inc., Alexandria, Virginia, 22303, USA.\\
arutth@rediffmail.com}

\author{A. K. Rajagopal}

\address{Inspire Institute Inc., Alexandria, Virginia, 22303, USA.}

\author{Sudha}

\address{Department of Physics, Kuvempu University, Shankaraghatta, Shimoga-577 451, India.\\ 
Inspire Institute Inc., Alexandria, Virginia, 22303, USA.}

\maketitle

\begin{history}
\received{20 May 2011}
\revised{18 June 2011}
\accepted{22 June 2011}
\end{history}

\begin{abstract}
Generalized measurement schemes on one part of  bipartite states, which would leave the set of {\em all}  separable states insensitive are explored here to understand {\em quantumness of correlations} in a more general perspecitve. This is done by employing linear maps associated with generalized projective measurements. A generalized measurement corresponds to a quantum operation mapping a density matrix to another density matrix, preserving its positivity, hermiticity and traceclass. The  Positive Operator Valued Measure (POVM) --  employed  earlier in the literature to optimize the measures of classical/quatnum correlations  -- correspond to  completely positive (CP) maps. The other class, the not completely positive (NCP) maps, are investigated here, in the context of measurements, for the first time. It is shown that  such NCP projective maps provide a new clue to the understanding of quantumness of correlations in a general setting. Especially, the separability-classicality dichotomy gets resolved only when both the classes  of projective maps (CP and NCP) are  incorporated as optimizing measurements. An explicit example of a separable state -- exhibiting non-zero quantum discord, when possible optimizing measurements are restricted to POVMs --  is re-examined  with this extended scheme incorporating NCP projective maps to elucidate the power of this approach.
\end{abstract}


\section{Introduction}	

 Entanglement between subsystems of a composite state brought forth perplexing distinctions~\cite{EPR} between classical and quantum correlations. Fundamental significance of such  incompatibility was highlighted by  Bell's novel work~\cite{Bell}. Following Werner~\cite{werner}, it is believed that  the statistical correlations between parts of  a convex mixture of product (separable) states can be reproduced by a classical hidden 
variable model and they satisfy all Bell inequalities.  The physical source of separable correlations  being a classical preparation device, they are termed classical. In other words, quantum correlation owes its origin to the impossibility of expressing a composite quantum state as a convex combination of product states.  However, several other measures of non-classical correlations --  which are more general than entanglement -- are drawing significant attention during the past few years~\cite{OZ,Ved,RR,Hor1,Partovi,Hamieh,BG,Luo,ARU,Wu,Modi1,Modi2,brub}. It is identified that  non-classical correlations, other than entanglement, offer quantum advantage in some information processing tasks~\cite{Datta,Piani}. 

Now we proceed to elaborate on the concept of quantumness of correlations -- other than that  implied by entanglement.  
In classical probability theory, two random variables $A$ and $B$
are said to be correlated if their  probability distribution, $P(a,b)$ 
cannot be expressed as a mere product of the marginal probabilities  
$P(a)$ and $P(b)$. Shannon mutual information 
\begin{eqnarray}
\label{shn}
H(A:B)&=&H(A)+H(B)-H(A,B)
\end{eqnarray}
(where $H(A,B)=-\sum_{a,b}\, p(a,b)\log p(a,b),\ H(A)=-\sum_{a}\, p(a)\log p(a)$, $H(B)=-\sum_{b}\, p(b)\log p(b)$) 
is an unequivocal measure of classical correlations. 

In the quantum description, probability distributions are replaced by  density operators and a bipartite density matrix $\hat\rho_{AB}$ is {\em correlated} if it cannot be expressed  in a simple  product form of its constituent subsystem density matrices $\hat\rho_A,\ \hat\rho_B$. 
The von Neumann mutual information,  
\begin{eqnarray}
\label{vN}
S(A:B)&=&S(\hat\rho_{AB}\vert\vert \hat\rho_A\otimes 
\hat\rho_B)\nonumber \\ 
&=&S(\hat\rho_{A})+S(\hat\rho_{B})-S(\hat\rho_{AB}) 
\end{eqnarray}
(where $S(\hat\rho)=-{\rm Tr}[\hat\rho\log \hat\rho]$) 
 quantifies the {\em total} correlations -- classical as well as quantum -- in a bipartite state $\hat\rho_{AB}$. Distinguishing these two
kinds of correlations gains basic importance -- that too when one addresses the issue from a significantly different perspective  -- keeping aside the established  separability-entanglement demarkation of correlations.  
It is with this view that  Ollivier and Zurek~(OZ)~\cite{OZ} pointed  towards  characterizing quantumness of correlations in a bipartite system  based on  measurement perspective. They  considered the quantum anologue of mutual information, which is sensitive to measurement on one part of the composite system as, 
\begin{equation}
\label{ozm}
{\cal S}(A:B)=S(\hat{\rho}_B)-\sum_{\alpha}\,p_\alpha\, S(\hat{\rho}_{B\vert A_\alpha})
\end{equation} 
where 
\begin{eqnarray}
\label{condiB}
\hat{\rho}_{B\vert A_\alpha}&=&\frac{\hat\Pi^A_\alpha\otimes I_B\, \hat\rho_{AB}\, \hat\Pi^A_\alpha\otimes I_B}{p_\alpha}\\
&=&\hat\Pi^A_\alpha\otimes \hat\rho^{B}_\alpha \nonumber  
\end{eqnarray}
denotes the conditional density operator, which results after a projective measurement  $\{\hat\Pi^A_\alpha\}$ on subsystem $A$ of the composite state 
$\hat\rho_{AB}$;\ \  $p_\alpha={\rm Tr}[\hat\Pi^A_\alpha\otimes I_B\, \hat\rho_{AB}]$ denotes the probability of outcome and $\hat\rho^{B}_\alpha={\rm Tr}_A[\hat{\rho}_{B\vert A_\alpha}]$. 

OZ proposed {\em quantum discord} as the {\em minimum} difference between the two equivalent quantum analogs (\ref{vN}) and (\ref{ozm}) of mutual information to characterize quantumness of correlations in $\hat{\rho}_{AB}$:   
 \begin{eqnarray}
 \label{discord}
 \delta(A,B)_{\{\hat\Pi^A_\alpha\}}&=&S(A:B)-{\rm max}_{\{\hat\Pi^A_\alpha\}}\, {\cal S}(A:B)
 \end{eqnarray} 
 where the maximization is done over complete, orthogonal projective measurements $\{\hat\Pi^A_\alpha\}$ on subsystem $A$. 

A classically correlated bipartite state remains insensitive to a specific choice of projective measurement  $\{\hat\Pi^A_\alpha\}$ on a part of the system   -- leading to vanishing quantum discord:    
\begin{eqnarray}
\delta(A,B)_{\{\hat\Pi^A_\alpha\}}&=&0\Rightarrow  \nonumber \\ 
\hat\rho^{\rm (cl)}_{AB}&=& 
\sum_\alpha \hat\Pi^A_\alpha\otimes I_B\, \hat\rho^{\rm (cl)}_{AB}\, \hat\Pi^A_\alpha\otimes I_B.
\end{eqnarray} 
Non-zero values of quantum discord quantify {\em quantumness} of correlations.  

Expressing the classically correlated state in the basis $\{\vert\alpha\rangle\}$  of the orthogonal projectors, it is easy to see that 
\begin{eqnarray}
\label{sepcl}
\hat\rho^{\rm (cl)}_{AB}&=& 
\sum_{\alpha,\beta',\beta}\,\langle \alpha;\beta' \vert\hat\rho^{\rm (cl)}_{AB}\vert \alpha;\beta\rangle \, \hat\Pi^A_\alpha \otimes \vert\beta'\rangle\langle \beta\vert\, \nonumber \\
&=&\sum_{\alpha}\,q_\alpha \, \hat\Pi^A_\alpha \otimes \hat\tau^B_\alpha 
\end{eqnarray}  
where $q_\alpha=\sum_{\beta}\, \langle \alpha;\beta \vert\hat\rho^{\rm (cl)}_{AB}\vert \alpha;\beta\rangle={\rm Tr}[\hat\rho^{\rm (cl)}_{AB}]$ and $\hat\tau^B_\alpha=\displaystyle\sum_{\beta,\beta'}\, \frac{\langle \alpha;\beta' \vert\hat\rho^{\rm (cl)}_{AB}\vert \alpha;\beta\rangle}{q_\alpha}\, 
\vert\beta'\rangle\langle \beta\vert$. Clearly, the classically correlated states form a subset of separable states of the form 
$\left\{\sum_{\alpha}\,q_\alpha \, \hat\Pi^A_\alpha \otimes \hat\tau^B_\alpha\right\}$. 
Quantum discord does not necessarily vanish for {\em all} separable states. In other words, it suggests that the concept of quantum correlations is more general than entanglement -- as separable states too exhibit quantumness of correlations (non-zero quantum discord).  OZ, however, based their 
 discussion on  quantumness of correlations by confining their attention only to orthogonal projective measurements $\{\hat\Pi^A_\alpha\otimes I_B\}$. 

We give here an example of a two qubit separable state, which has non-zero quantum discord~\cite{Ved,Hamieh}:  
\begin{eqnarray}
\label{hamieh} 
\hat\rho_{AB}&=&p\, \vert 0_A,0_B\rangle\, \langle 0_A,0_B\vert + (1-p)\,  \vert +_A,+_B\rangle\, \langle 
+_A,+_B\vert,\nonumber \\ 
&& \ \ \ \ \ 0\leq p\leq 1,\ \ \vert \pm\rangle = \frac{1}{\sqrt{2}}\,(\vert 0\rangle\pm\vert 1\rangle). 
\end{eqnarray}

Following a similar approach Henderson and
Vedral (HV)~\cite{Ved}  independently investigated how to separate classical and quantum
correlations. They employed general positive operator valued measures (POVMs) to quantify classical correlations in the state 
$\hat{\rho}_{AB}$ in terms of the residual information entropy of $B$ as follows: 
 \begin{equation}
 C_A(\hat\rho_{AB})={\rm max}_{\{V^A_i\}}\, S(\hat\rho_B)-\sum_i\, q_i S(\hat\rho^B_i) 
 \end{equation} 
where  $\hat\rho^B_i={\rm Tr}_{A}[V^A_i\otimes I_B\, \hat\rho_{AB}\, 
V_i^{A\dag}\otimes I_B]/q_i$ is the density matrix of subsystem $B$ after the measurement $\{V^A_i\otimes I_B\}$ is performed on $A$ and $q_i={\rm Tr}_{AB}[V^A_i\otimes I_B\, \hat\rho_{AB}\, V_i^{A\dag}\otimes I_B]$ denotes the probability of outcome. 
In a classically correlated state the residual information entropy of $B$ {\em does not} increase under an optimal measurement scheme on $A$.    

By analyzing some examples HV found that classical and entangled correlations do not add up to give  
total correlations~\cite{Ved} i.e., $C_A(\hat\rho_{AB})+E_{\rm RE}(\hat\rho_{AB})\neq S(A:B).$, where $E_{\rm RE}(\hat\rho_{AB})$ denotes the relative entropy of entanglement.   Hamieh et. al.~\cite{Hamieh} showed that 
optimization of classical correlations~\cite{Ved} in two qubit states may be achieved 
using orthogonal projective measurements themselves. This also leads to the identification that  the classical 
correlations~\cite{Ved} and the quantum discord~\cite{OZ} add up to give the mutual information entropy in two-qubit states.

Another measure of quantum correlations  is the one-way information
deficit~\cite{Hor1} which is defined as the
minimal increase of entropy after a projective  measurement $\{\hat\Pi^{A}_\alpha\}$
on subsystem $A$ is done:
\begin{equation}
\Delta^\rightarrow(\hat{\rho}_{AB})={\rm min}_{\{\hat\Pi^{A}_\alpha\}}\, S\left(\sum_\alpha\,\hat\Pi^{A}_\alpha\, \hat{\rho}_{AB}\, 
\hat\Pi^{A}_\alpha\right)-S(\hat{\rho}_{AB}).
\end{equation}
 The one-way information deficit vanishes
  only on states with zero quantum discord.  

Quantum discord $\delta(A,B)_{\{\hat\Pi^A_\alpha\}}$, the HV classical correlations $C_A(\hat\rho_{AB})$ and the one-way information deficit $\Delta^\rightarrow(\hat{\rho}_{AB})$  are  all asymmetric with respect to measurements on the subsystems $A$ and $B$. Quantum deficit --  one other measure of non-classical correlations --  which is symmetric about the subsystems $A$, $B$, was proposed by Rajagopal and Rendell~\cite{RR} as follows: 
\begin{equation}
 \label{dab}
 D_{AB}=S(\hat\rho_{AB}\vert\vert\hat\rho^{(d)}_{AB})
 = {\rm Tr}\, [\hat\rho_{AB}\log\hat\rho_{AB}]-{\rm Tr}\, [\hat\rho_{AB}\log\hat\rho^{d}_{AB}],  
 \end{equation}
 where  $\hat\rho^{(d)}_{AB}=\sum_{a,b} P(a,b)\,\hat\Pi^A_a \otimes \hat\Pi^B_b$, where $\hat\Pi^A_a$, $\hat\Pi^B_b$ correspond to  eigenprojectors of the subsystems $\hat\rho_A,\ \hat\rho_B$ 
with $P(a,b)=\langle a,b\vert\hat\rho_{AB}\vert a,b \rangle$ denoting the diagonal elements of $\hat\rho_{AB}$, in its subsystem eigen basis and  $P(a)=\sum_{b} P(a,b),\ P(b)=\sum_{a} P(b,a)$ the eigenvalues of $\hat\rho_A,\ \hat\rho_B$ respectively. The quantum deficit $D_{AB}$ determines the quantum excess of correlations in the state $\hat\rho_{AB}$, with reference to its classically decohered counterpart  $\hat\rho^{(d)}_{AB}$ and it  
 vanishes {\em iff} $\hat\rho_{AB}\equiv \hat\rho^{(d)}_{AB})$. It may be noted that  bipartite states with zero quantum deficit have vanishing quantum discord. Another important feature is that evaluating quantum deficit is easier compared to the other measures of correlations outlined above, as no optimization procedure is involved in its evaluation.

It appears natural to raise the question~\cite{ARU}: are there more general measurement schemes on one part of  bipartite states, which would leave {\em all} the separable states insensitive? Possibility of such generalized measurements would resolve the dichotomy of separability vs classicality of correlations. Furthermore any measure of quantumness of correlations could then be identified with that of entanglement itself. In this paper, we analyze the basic features of generalized  measurement scheme which could imply that absense of entanglement and classicality are synonymous. We show that not completely positive (NCP) projective maps  -- in contrast to POVMs -- are the essential ingredients of  generalized measurements on one end of a bipartite system that leave separable states unaltered.   

\section{A generalized measure of quantumness of correlations}      

We discuss some specific properties of quantum discord so as to  extend the notion of {\em quantumness of correlations} in a bipartite system  by invoking generalized measurements.   

Consider a bipartite state $\hat\rho_{AB}$,  for which 
optimization of  quantum discord $\delta(A,B)_{\{\hat\Pi^A_\alpha\}}$ is realized in terms of a complete orthogonal projective set $\{\hat\Pi^A_\alpha\}$. 
The  state left  after measurement is given by,
\begin{eqnarray}   
\hat\rho^{\cal D}_{AB}&=&\sum_\alpha\,  \hat\Pi_{\alpha}^{A}\otimes I_B  
\hat\rho_{AB}  \hat\Pi_{\alpha}^{A}\otimes I_B \nonumber \\ 
&=&\sum_\alpha\, p_\alpha\, \hat{\rho}_{B\vert A_\alpha}
\end{eqnarray}
where $\hat{\rho}_{B\vert A_\alpha}$ is the conditional density operator (see (\ref{condiB})) and $p_\alpha={\rm Tr}[\hat\Pi^A_\alpha\otimes I_B\, \hat\rho_{AB}]$. Using the property~\cite{OZ}  
\begin{eqnarray}
S(\hat\rho^{\cal D}_{AB})=-\sum_\alpha p_\alpha\log p_\alpha +\sum_{\alpha} p_\alpha S(\hat{\rho}_{B\vert A_\alpha})\nonumber \\ 
\end{eqnarray} 
one can express  quantum discord (see (\ref{discord})) in terms of the relative entropies as follows:  
\begin{eqnarray} 
\label{discord2}
 \delta(A,B)_{\{\hat\Pi^A_\alpha\}}&=&S(\hat\rho^{\cal D}_{AB})-S(\hat\rho_{AB})+S(\hat\rho_A)-\sum_\alpha p_\alpha\log p_\alpha \nonumber \\
 &=& S(\hat\rho_{AB}\vert\vert \hat\rho^{\cal D}_{AB}) + S(\hat\rho_{A}\vert\vert \hat\rho^{\cal D}_{A})
\end{eqnarray}
This structure of quantum discord clearly projects  out the fact that (i) $\delta(A,B)_{\{\hat\Pi^A_\alpha\}}\geq 0$ as the relative entropies $S(\hat\rho_{AB}\vert\vert \hat\rho^{\cal D}_{AB}),\ S(\hat\rho_{A}\vert\vert \hat\rho^{\cal D}_{A})$ are positive semidefinite quantities (ii) they vanish iff  $\hat\rho_{AB}\equiv \hat\rho^{\cal D}_{AB}$ i.e., if the state $\hat\rho_{AB}$ remains insensitive to projective measurement $\{\hat\Pi^A_\alpha\}$. 
Moreover, observing that the state after measurement is a classically correlated state i.e., $\hat\rho^{\cal D}_{AB}=\sum_\alpha\, p_\alpha\ \hat\Pi^A_\alpha \otimes \rho^{B}_\alpha$, the quantum discord gets related to~\cite{Modi1} {\em  distance between the given state $\hat{\rho}_{AB}$ and its closest classically correlated state} $\hat\rho^{\cal D}_{AB}$ (where {\em distance} is considered in terms of the relative entropy). 

A natural extension of the measure of quantumness of correlations~\cite{ARU} -- as a distance between the given bipartite state and the closest state realized after measurements at one end of the state -- will be outlined in the following.

Let us consider the set of all tripartite   density operators $\{\hat\rho_{A'AB}\}$ in an extended Hilbert space 
${\cal H}_{A'}\otimes{\cal H}_A\otimes{\cal H}_B$, such that the bipartite 
state $\hat\rho_{AB}$ under investigation is a marginal of this extended system:    
\begin{equation} 
{\rm Tr}_{A'}[\hat\rho_{A'AB}]=\hat\rho_{AB}.
\end{equation} 
Now, carrying out an orthogonal projective measurement $\Pi_{i}^{(A'A)}; i=1,2,\ldots , $ 
on one of the subsystems $A'A$ of the tripartite state $\hat\rho_{A'AB}$ we obtain,      
 
\begin{eqnarray}
\hat\rho_{A'AB}&\rightarrow &\hat\rho^{(i)}_{A'AB}=\frac{1}{p_i}\,\left[\hat\Pi^{(A'A)}_i\otimes I_B\, 
\hat\rho_{A'AB}\, 
\hat\Pi^{(A'A)}_i\otimes I_B\right]  \\
{\rm and \ \ }\hat\rho_{AB}&\rightarrow &\hat\rho^{(i)}_{AB}=\frac{1}{p_i}\,{\rm 
Tr}_{A'}\left[\hat\Pi^{(A'A)}_i\otimes I_B\, 
\hat\rho_{A'AB}\, \hat\Pi^{(A'A)}_i\otimes I_B\right]  \nonumber
\end{eqnarray}
where $p_i~=~{\rm Tr}_{A'AB}~[\hat\Pi^{(A'A)}_i~\otimes~I_B~\hat\rho_{A'AB}]$ denotes the  
probability of occurrence of $i^{\rm th}$ outcome.   

We define {\em Quantumness} ${\cal Q}_{AB}$ associated with a bipartite state 
$\hat\rho_{AB}$  as the  relative entropy 
\begin{eqnarray}
\label{qab}
&{\cal Q}_{AB}={\rm min}_{\{\hat\Pi^{(A'A)}_i,\, \hat\rho_{A'AB}\}}\,   S(\hat\rho_{AB}\vert\vert 
\hat\rho^{\cal R}_{AB})
\end{eqnarray} 
Here, 
$\hat\rho^{\cal R}_{AB}={\rm Tr}_{A'}\,[\sum_i\, \hat\Pi^{(A'A)}_i\otimes I_B\, 
\hat\rho_{A'AB}\, \hat\Pi^{(A'A)}_i\otimes I_B],$
  denotes the residual state of the bipartite system, left after the generalized projective measurement 
is performed.  The minimum in Eq.~(\ref{qab}) is taken over the set $\{\hat\Pi^{(A'A)}_i\}$ of  projectors
 on the subsystems $A'A$ of {\em all} possible extendend states  $\{\hat\rho_{A'AB}\}$, which contain 
 the given bipartite state $\hat\rho_{AB}$ as their marginal system.  
 
The quantumness, ${\cal Q}_{AB}\geq 0$ (by definition), for all generalized measurements  - the equality sign
holding iff  $\hat\rho^{\cal R}_{AB}=\hat\rho_{AB}$ i.e.,  quantumness vanishes iff the bipartite state $\hat\rho_{AB}$ 
 remains insensitive to generalized measurement  $\{\hat\Pi^{(A'A)}_i\}$. 

Corresponding to a chosen measurement scheme $\{\hat\Pi_{i}^{A'A}\}$  
 we may express the extended state $\hat\rho_{A'AB}$ in terms of the complete, orthogonal set of basis states 
$\{\vert i\rangle_{A'A} \otimes \vert \beta\rangle_{B}\}$  as, 
\begin{equation}
\hat\rho_{A'AB}=\displaystyle\sum_{i',i, \beta',\beta}\, P(i',\beta'; i,\beta)\, 
\vert i'\rangle_{A'A}\, \langle i'\vert  \otimes \vert \beta'\rangle_{B}\langle \beta\vert. \nonumber \\  
\end{equation}
We then obtain,   
\begin{equation}
\label{rho'}
\hat\rho^{\cal R}_{A'AB}=\hat\Pi^{(A'A)}_i\otimes I_B\, \hat\rho_{A'AB}\, 
\hat\Pi^{(A'A)}_i\otimes I_B=\displaystyle\sum_{\beta',\beta}\, P(i,\beta'; i,\beta)\, \hat\Pi^{(A'A)}_i\otimes 
\vert\beta'\rangle_B \langle\beta\vert  
\end{equation}
which leads in turn to 
\begin{eqnarray}
\label{sep2}
\hat\rho^{\cal R}_{AB}&=& {\rm Tr}_{A'}\, \left[\displaystyle\sum_i\, 
\hat\Pi^{(A'A)}_i\otimes I_B\, \hat\rho_{A'AB}\, 
\hat\Pi^{(A'A)}_i\otimes I_B\right] \nonumber \\
& =& \displaystyle\sum_i\, p_i\, \hat\rho^{A}_i\otimes \hat\rho^{B}_{i} 
\end{eqnarray}
where $\hat\rho^{A}_i={\rm Tr}_{A'}\,[\hat\Pi^{(A'A)}_i]$
and   
\begin{eqnarray*}
\hat\rho^{B}_i&=&\sum_{\beta',\beta}\,\frac{P(i,\beta';i,\beta)}{p_i}\, 
\vert\beta'\rangle_B\langle\beta\vert, \\ 
p_i&=&{\rm Tr}\,  [\hat\Pi^{(A'A)}_i\otimes I_B \hat\rho_{A'AB}]=\sum_{\beta}\,P(i,\beta;i,\beta). 
\end{eqnarray*}

Clearly, the state $\hat\rho^{\cal R}_{AB}$ of the bipartite system --  left 
after performing the generalized measurement $\{\hat\Pi^{(A'A)}_i\}$ on  the part $A'A$ of the global system --    
 is a separable state. As the optimization of quantumness ${\cal Q}_{AB}$ is done over the set of 
all projectors $\{\hat\Pi^{(A'A)}_i\},$ and the set of  all   extended states $\{\hat\rho_{A'AB}\}$, it is readily seen 
that   $\{\hat\rho^{\cal R}_{AB}=\hat\varrho^{{\rm (sep)}}_{AB}; \hat\rho_B={\rm Tr}[\hat\rho^{\cal R}_{AB}] \}$
 corresponds to  the set of  all separable states which share the same subsystem density matrix $\hat\rho_B$ for the 
part $B$ (i.e., the subsystem, which does not come under the direct action of generalized measurements $\{\hat\Pi^{(A'A)}_i\}$). We thus obtain 
\begin{eqnarray}
\label{relent}
{\cal Q}_{AB}&=&{\rm min}_{\{\hat\Pi^{(A'A)}_i,\, \hat\rho_{A'AB}\}}\,   S(\hat\rho_{AB}\vert\vert 
\hat\rho^{\cal R}_{AB})\nonumber \\ 
&=& {\rm min}_{\{\hat\varrho_{AB}^{{\rm (sep)}}\}}\, S(\hat\rho_{AB}\vert\vert \hat\varrho^{{\rm (sep)}}_{AB})
\end{eqnarray}
with minimization taken over the set of all separable states $\{\hat\varrho^{{\rm (sep)}}_{AB}; \hat\rho_B={\rm 
Tr}[\hat\varrho^{\rm (sep)}_{AB}]\}.$ 

In other words, the generalized measure ${\cal Q}_{AB}$ of quantumness of correlations    
corresponds to the distance between the given  state $\hat\rho_{AB}$  with the closest separable state 
$\hat\varrho^{{\rm (sep)}}_{AB};\ {\rm Tr}\,[\hat\varrho^{{\rm (sep)}}_{AB}]=\hat\rho_B.$  
From Eq.~(\ref{relent}) it is evident that quantumness ${\cal Q}_{AB}$ is 
necessarily non-zero for all entangled bipartite states $\hat\rho_{AB}$ and vanishes for {\em all} separable states.  
Moreover, ${\cal Q}_{AB}$ also serves as an upper bound to   
the relative entropy of entanglement~\cite{PlVed2}. While the evaluation of  ${\cal Q}_{AB}$ is as hard a task as that of relative entropy of entanglement, the significant point here is that it brings out the required generalized scheme of measurements, which resolve the dichotomy between 
{\em quantumness of correlations} and {\em entanglement}. Further, the established connection   -- viz., the quantumness of correlations is the  distance between the given bipartite state  with its closest separable state ( sharing the same marginal state for the subsystem $B$) --  highlights the merger of quantumness of correlations with quantum entanglement itself. This in turn ensures that any other {\em operational} measure of bipartite entanglement would faithfully reflect quantumness of correlations in the state.

We illustrate the scheme of generalized projective measurements on $A'A$ subsystem of an extended tripartite state $\hat\rho_{A'AB}$ of the separable state ({\ref{hamieh}). An extended  three qubit state 
\begin{eqnarray}
\label{a'ab}
\hat\rho_{A'AB}&=&p\, \vert 1_{A'},0_A,0_B\rangle\, \langle 1_{A'}, 0_A,0_B\vert 
 + (1-p)\,  
\vert 0_{A'}, +_A,+_B\rangle\, \langle 0_{A'},+_A,+_B\vert
\end{eqnarray} 
leads to the given two qubit state (\ref{hamieh}) by tracing over the $A'$ qubit. We find that  
the complete, orthogonal set of projectors $\{\hat\Pi_{i}^{(A'A)}\}$ on $A'A$ constituted by  
\begin{eqnarray}
\label{gmp}
\hat\Pi_{1}^{(A'A)}&=&\vert 0_{A'},+_A\rangle\, \langle 0_{A'},+_A\vert,  \nonumber \\
\hat\Pi_{2}^{(A'A)}&=&\vert 0_{A'},-_A\rangle\, \langle 0_{A'},-_A\vert,\nonumber \\ 
\hat\Pi_{3}^{(A'A)}&=&\vert 1_{A'},0_A\rangle\, \langle 1_{A'},0_A\vert, \nonumber \\ 
\hat\Pi_{4}^{(A'A)}&=&\vert 1_{A'},1_A\rangle\, \langle 1_{A'},1_A\vert 
\end{eqnarray}
 leaves the overall state (\ref{hamieh})  unaltered:
 \begin{equation} 
 \label{insensitive}
\hat\rho^{\cal R}_{A'AB}=\sum_{i=1}^{4}\, \hat\Pi^{(A'A)}_i~\otimes~I_B\, \hat\rho_{A'AB}\, \hat\Pi^{(A'A)}_i~\otimes~I_B\equiv\hat\rho_{A'AB}.
\end{equation} 
So, we identify that the bipartite state (\ref{hamieh}) is insensitive under the generalized projective measurements (\ref{gmp}))   i.e.,   $\hat\rho^{\cal R}_{AB}=\hat\rho_{AB}$ implying that ${\cal Q}_{AB}=S(\hat\rho_{AB}\vert\vert\hat\rho^{\cal R}_{AB})=0$ in this state.   

The generalized projective measurements on  $A'A$ part of the extended state   may be viewed as quantum maps, which transform  density matrices $\hat\rho_A$ (before measurement) to density matrices $\hat\rho^{\cal R}_A$ (after measurement) -- preserving their hermiticity, positivity and trace class.  In the next section we investigate the properties of the linear map associated with the generalized measurements. 

\section{Linear ${\cal A},{\cal B}$ maps associated with generalized projective measurements} 
Dynamical ${\cal A}$ and ${\cal B}$ maps have been employed extensively by Sudarshan and co-workers to investigate open system evolution of quantum systems~\cite{ECGS,Jordan1,Jordan}. Here, we elucidate the projective measurements $\{\hat\Pi_{i}^{(A'A)}\}$ on  $\hat{\rho}_{A'A}$ in terms of    linear ${\cal A}, {\cal B}$ quantum maps  on $\hat\rho_A$  -- transforming it  to the resultant density matrix $\hat\rho^{\cal R}_A$  -- preserving the positivity, hermiticity and unit trace conditions. The elements  $\left(\hat\rho^{\cal R}_A\right)_{a_k a_l}$  after measurement are explicitly expressed in terms of those of  initial density matrix  $\left(\hat\rho_A\right)_{a_i a_j}$ via the ${\cal A}$ map  as~\cite{ECGS,Jordan1} 
\begin{equation}
\label{defA}
\left(\hat\rho_A^{\cal R}\right)_{a_i a_j}=\sum_{a_k, a_l}\, {\cal A}_{a_ia_j;a_k a_l}\, \left(\hat\rho_A\right)_{a_k a_l}.
\end{equation} 
That the resultant density matrix $\hat\rho^{\cal R}_A$  is Hermitian and has unit trace  leads to the conditions 
\begin{eqnarray}
\label{A1}
{\rm Hermiticity}:&&\ \ {\cal A}_{a_ia_j;a_k a_l}={\cal A}^*_{a_ja_i;a_la_k }, \\
\label{A2}
{\rm Trace\ preservation}:&&\ \ \sum_{a_i} {\cal A}_{a_ia_i;a_k a_l}=\delta_{a_k,a_l}, 
\end{eqnarray}
In order to bring out the properties (\ref{A1}),(\ref{A2}) in a lucid manner,  a realigned matrix ${\cal B}$  ~\cite{ECGS,Jordan1}:
\begin{equation}
\label{abdef}
{\cal B}_{a_i a_k;a_j a_l}={\cal A}_{a_i a_j;a_k a_l}. 
\end{equation}  
The hermiticity property (\ref{A1}) leads to the condition ${\cal B}_{a_ia_k;a_j a_l}={\cal B}^*_{a_ja_l;a_ia_k}$, i.e., the  map ${\cal B}$ is hermitian.

In terms of the spectral decomposition ${\cal B}_{a_i a_k;a_j a_l}=
\sum_{\alpha}\, \lambda_\alpha M^{(\alpha)}_{a_i a_k} M^{(\alpha)*}_{a_j a_l }$,  the action of the ${\cal B}$ map on the density matrix is then readily identified as,  
\begin{eqnarray} 
\left(\hat\rho^{\cal R}_{A}\right)_{a_i a_j}&=& \sum_{\alpha, a_k,a_l}\, \lambda_\alpha M^{(\alpha)}_{a_i a_k} M^{(\alpha)*}_{a_j a_l }\, \left(\hat\rho^{\cal R}_A\right)_{a_k a_l}\nonumber \\ 
&&\Rightarrow \ \hat\rho^{\cal R}_A = \sum_{\alpha}\, \lambda_\alpha\, M^{(\alpha)}\, \hat\rho_A\, M^{(\alpha)\dag} 
\end{eqnarray} 
and this corresponds to POVM on $\hat{\rho}_A$ provided $\lambda_\alpha\geq 0$ or a completely positive (CP) map associated with projective measurement; 
otherwise it is a not completely positive (NCP) map. 

We focus on finding the CP/NCP nature of the projective quantum map transforming the single qubit state 
$\hat\rho_A={\rm Tr}_{A'B}[\hat\rho_{A'AB}]$ (before measurement) with $\hat{\rho}^{\cal R}_A={\rm Tr}_{A'B}\left[\left\{\sum_i\Pi^{(A'A)}_i\hat\rho_{A'AB}\Pi^{(A'A)}_i\right\}\right]$ (after measurement) -- corresponding to the specific measurement  scheme $\{\Pi^{(A'A)}_i\}$ (see Eq.~(\ref{gmp}))  on the state $\hat\rho_{A'AA}$ of  Eq.~(\ref{a'ab}) -- i.e., in the specific example discussed in Sec.~2. 
It is pertinent to point out here  that the state $\hat\rho_{A'AB}$, and hence the reduced state $\hat\rho_A$, remain insensitive to the projective measurement (\ref{gmp}), as has already been illustrated  explicitly in Sec.~2 (see Eqs.(\ref{a'ab}-\ref{insensitive})). The corresponding quantum map transforming $\hat\rho_A\longrightarrow \hat{\rho}_A^{\cal R}$ must reveal this insensitivity.

In order to deduce the explicit structure of the projective {\cal A}, {\cal B} maps, we employ the concept of  {\em assignment map}~\cite{RRK}. Explicit technical details and derivations are elaborated in Appendix.  We obtain  the ${\cal B}$ map (see Appendix Eq.~(\ref{31})) associated with this particular example as, 
\begin{equation}
\label{ncpb}
{\cal B} = \left(\begin{array}{cccc} 1 & 0 & 0 & \frac{1}{2} \\ 
0 & 0 & \frac{1}{2} & 0 \\
0 & \frac{1}{2} & 0 & 0  \\
\frac{1}{2} & 0 & 0 & 1 
\end{array}\right)
\end{equation} 
where the rows and columns are labeled as  $\{00, 01,10, 11\}$. The associated ${\cal A}$ matrix  is then  obtained as ( using (\ref{abdef})), 
\begin{equation}
\label{ncpa}
{\cal A} = \left(\begin{array}{cccc} 1 & 0 & 0 & 0 \\ 
0 & \frac{1}{2} & \frac{1}{2} & 0 \\
0 & \frac{1}{2} & \frac{1}{2} & 0  \\
0 & 0 & 0 & 1 
\end{array}\right)
\end{equation} 
\begin{itemize} 
\item Applying  the measurement map ${\cal A}$ of Eq.~(\ref{ncpa}) on the state $\hat\rho_A={\rm Tr}_{A'B}[\hat\rho_{A'AB}]=p\vert 0_A\rangle\langle 0_A\vert+(1-p)\, \vert +_A\rangle\langle+_A\vert$ (the state before measurement)   it may be seen explicitly (following Eq.~(\ref{defA}))  that   $\left(\hat\rho_A^{\cal R}\right)_{a_ia_j}=\displaystyle\sum_{a_k,a_l=0,1}{\cal A}_{a_ia_j;a_ka_l}\, \left(\hat\rho_A\right)_{a_ka_l}\equiv [\hat\rho_A]_{a_ia_j}$ i.e., the state is insensitive to this measurement.  It may be recalled here that the projective measurement (\ref{gmp})  leaves the tripartite state (\ref{a'ab}) -- and hence its subsystems $\hat\rho_{AB}$  (and also $\hat\rho_A$)  -- undisturbed as is illustrated in Sec.~2. This in turn led to the implication that the quantumness of correlation  ${\cal Q}_{AB}$ vanishes for the  separable state $\hat\rho_{AB}$ of Eq.~(\ref{hamieh})  -- whereas, quantum discord and quantum deficit are  non-zero.    
\item The eigenvalues of ${\cal B}$ are readily found to be $\left(\frac{1}{2},\frac{1}{2},-\frac{1}{2},\frac{3}{2}\right)$,  implying that the projective measurement (\ref{gmp}) on the state $\hat\rho_A$ corresponds to a NCP map.
\end{itemize} 

In other words, we reach a crucial identification that the map, which leaves the state $\hat{\rho}_{AB}$ of Eq.~(\ref{hamieh}) insensitive under measurements is  NCP. Our generalized measure of quantumness (\ref{qab}) may also be expressed as, 
\begin{eqnarray}
\label{qcpncp}
{\cal Q}_{AB}={\rm min}_{\{{\rm CP/NCP\ projective\ maps\ on}\ A\}}\,   S(\hat\rho_{AB}\vert\vert 
\hat\rho^{\cal R}_{AB})
\end{eqnarray} 
where we emphasize that positivity, hermiticity and trace of the given density matrix are preserved by the optimizing CP/NCP projective maps. 
A comparison of Eq.~(\ref{qcpncp}) with the alternate form   (given in Eq.~(\ref{relent})), suggests that both the classes  of projective maps (CP and NCP) need to be incorporated in order to deem  quantumness of correlations as synonymous with quantum entanglement itself.  
Having thus established that the quantumness of correlations ${\cal Q}_{AB}$ of bipartite states is non-zero only for entangled states, we  point out once again that any other {\em operational} measure of entanglement would necessarily imply such non-classicality of correlations --  and this identification takes away the burden of  evaluating ${\cal Q}_{AB}$ (where the optimization procedure turns out to be a demanding task) per se to infer quantumness.      

\section{Summary}
Sudarshan and coworkers~\cite{ECGS,Jordan1} put forward the conceptual formulation of  quantum theory of open system evolution in terms of dynamical ${\cal A}, {\cal B}$ maps almost 50 years ago and they also investigated it in the more general setting~\cite{Jordan,Rosario,Modi3} -- including NCP dynamical maps. In this paper we  highlight the important role of  NCP projective maps in the context of  measurements. It is shown  that incorporating generalized measurement schemes -- including both CP  as well as NCP  maps -- resolves the dichotomy of separability vs classicality of correlations.  

\section*{Acknowledgments}

We thank insightful discussions with Dr. Kavan Modi on assignment maps.

\section*{Appendix: ${\cal A}$, ${\cal B}$ maps associated with projective measurement}

Let us consider complete, orthonormal set of projective measurements $\{\Pi^{(A'A)}_i\}$ on  $\hat\rho_{A'A}$. 
We proceed to construct the ${\cal A},{\cal B}$ maps transforming the system state $\hat\rho_A={\rm Tr}_{A'}[\hat\rho_{A'A}]$ (before measurement) to the state $\hat\rho^{\cal R}_A={\rm Tr}_{A'}[\sum_i\, \hat\Pi^{A'A}_i\, \hat\rho_{A'A}\hat\Pi^{A'A}_i]$  (after measurement). 
\begin{eqnarray}
\label{9}
\hat\rho^{\cal R}_{A}&=&{\rm Tr}_{A'}\left[\sum_i\, \hat\Pi^{A'A}_i\, \hat\rho_{A'A}\, \hat\Pi^{A'A}_i\right]\nonumber \\
&=& \sum_i\, {\rm Tr}_{A'A}[\hat\Pi^{A'A}_i\, \hat\rho_{A'A}]\, {\rm Tr}_{A'}[\hat\Pi^{A'A}_i] \nonumber \\
&=&  \sum_i\, {\cal P}_i\ \hat\rho^{A}_{i}
\end{eqnarray}
where we have denoted
\begin{eqnarray}
\label{rhoai}
{\rm Tr}_{A'}[\hat\Pi^{A'A}_i]&=&\hat\rho^{A}_i \\
 {\cal P}_i&=&{\rm Tr}_{A'A}[\hat\Pi^{A'A}_i\, \hat\rho_{A'A}]
\end{eqnarray} 

We  simplify ${\cal P}_i={\rm Tr}_{A'A}[\hat\Pi^{A'A}_i\, \hat\rho_{A'A}]$ in order to construct the associated ${\cal A}$ map as follows:  
\begin{equation}
\label{11}
{\rm Tr}_{A'A}[\hat\Pi^{A'A}_i\, \hat\rho_{A'A}]={\rm Tr}_{A'A}[\hat\Pi^{A'A}_i\, {\cal A}(\hat\rho_A)]={\cal T}\circ \Pi\circ 
\tilde{\cal A}(\hat\rho_A),
\end{equation}   
where $\tilde{\cal A}(\hat\rho_{A})=\hat\rho_{A'A}$ defines the {\em assignment map}~\cite{RRK}. The assingment map is linear i.e.,   
\begin{eqnarray}
\label{12}
 \tilde{\cal A}(P^{A}_\alpha)&=&\tau^{A'}_\alpha\otimes P_\alpha^{A}\\
\Rightarrow \tilde{\cal A}\left(\sum_k r_\alpha\, P^{A}_\alpha\right)&=&\sum_\alpha\, r_\alpha\ \tau^{A'}_\alpha\otimes P^{A}_\alpha 
\end{eqnarray}
where $P^{A}_\alpha$ are  linearly independent states of system $A$.  
Let $\{Q_\beta\}$ be a set of hermitian operators such that 
\begin{eqnarray}
\label{13} 
{\rm Tr}[P^{A}_\alpha\, Q_\beta]&=&\delta_{\alpha,\beta}\nonumber \\
\sum_\beta\, Q_\beta=I_A.
\end{eqnarray}
We can thus express
\begin{eqnarray} 
\label{14}
\tilde{\cal A}=\sum_\alpha\ \tau_\alpha^{A'}\otimes P_\alpha^{A}\otimes Q_\alpha^{T}. 
\end{eqnarray}
(With the above construction, it may be readily identified that 
\begin{eqnarray*}
 \left[\tilde{\cal A}\left(\sum_\beta r_\beta\, P^{A}_\beta\right)\right]_{a'_ia_i;a'_ja_j}&=&\sum_\alpha\sum_{a_k,a_l}\ \left[\tau_\alpha^{A'}\otimes P_\alpha^{A}\otimes  Q_\alpha^{T}\right]_{a'_ia_ia_k;a'_ja_ja_l} 
 \left[\sum_\beta r_\beta\, P^{A}_\beta\right]_{a_k a_l}\nonumber \\
 &=&\sum_{\alpha,\beta}\ r_\beta\,\sum_{a_k,a_l}\ \left[\tau_\alpha^{A'}\otimes P_\alpha^{A}\right]_{a'_ia_i;a'_ja_j}\ \left[Q_\alpha^{T}\right]_{a_k a_l} 
 \left[  P^{A}_\beta)\right]_{a_k a_l}\nonumber \\ 
 &=& \sum_{\alpha,\beta}\ r_\beta\, \left[\tau_\alpha^{A'}\otimes P_\alpha^{A}\right]_{a'_ia_i;a'_ja_j}\ {\rm Tr}[P^{A}_\beta\, Q_\alpha]\nonumber \\ 
&=& \sum_{\alpha,\beta}\ r_\beta\, \left[\tau_\alpha^{A'}\otimes P_\alpha^{A}\right]_{a'_ia_i;a'_ja_j}\ \delta_{\alpha,\beta}\nonumber \\ 
&=& \sum_{\alpha}\ r_\alpha\, \left[\tau_\alpha^{A'}\otimes P_\alpha^{A}\right]_{a'_ia_i;a'_ja_j} 
\end{eqnarray*}
or we obtain, $\tilde{\cal A}\left(\sum_\alpha\, r_\alpha\, P^{A}_\alpha\right)= \sum_{\alpha}\ r_\alpha\, \tau_\alpha^{A'}\otimes P_\alpha^{A}$ as expected).
Substituting (\ref{14}) in (\ref{11}) we obtain,
\begin{eqnarray}
\label{15}
{\cal P}_i={\rm Tr}_{A'A}[\hat\Pi^{A'A}_i\, \hat\rho_{A'A}]&=&{\rm Tr}_{A'A}[\hat\Pi^{A'A}_i\, \tilde{\cal A}(\hat\rho_{A})]\nonumber \\
&=& \sum_{a'_i,a_j,a'_k,a_l,a_s,a_t}\left[\hat\Pi^{A'A}_i\right]_{a'_i,a_j;,a_l,a_s}\, \left[\tilde{\cal A}]\right]_{a'_ia_ja_s;a'_ka_la_t}\left[(\hat\rho_{A})\right]_{a_sa_t}\nonumber \\
\end{eqnarray}  
 and substituting (\ref{15}) back in (\ref{9}), we  identify the following:  
\begin{eqnarray}
 \left[\hat\rho^{\cal R}_A\right]_{a_u a_v}&=&\sum_{i} \left[\hat\rho^{A}_i\right]_{a_u a_v}\ \left\{
 \sum_{a'_i,a_j,a'_k,a_l,a_s,a_t}\left[\hat\Pi^{A'A}_i\right]_{a'_i,a_j;,a_l,a_s}\, \left[\tilde{\cal A}\right]_{a'_ia_ja_s;a'_ka_la_t}\, \left[(\hat\rho_{A})\right]_{a_sa_t}
 \right\}\nonumber \\ 
&=&\sum_{a_s,a_t} {\cal A}_{a_u a_v;a_sa_t}\, \left[(\hat\rho_{A})\right]_{a_sa_t} 
\end{eqnarray}
 where we identify the elements of the ${\cal A}$-matrix transforming the intital state $\rho_{A}$ to final state $\rho^{\cal R}_A$ as,   
 \begin{eqnarray}
  {\cal A}_{a_u a_v;a_sa_t}&=&\sum_{i} \left[\hat\rho^{A}_i\right]_{a_sa_t}  \left\{ \sum_{a'_i,a_j,a_l,a_s}\left[\hat\Pi^{A'A}_i\right]_{a'_i,a_j;,a_l,a_s}\, \tilde{\cal A}_{a'_ia_ja_s;a'_ka_la_t}
 \right\}\nonumber \\ 
  &=&\sum_{i}\, \sum_{a'_i,a_j,a_l,a_s} \left[\hat\rho^{A}_i\right]_{a_sa_t}\, \left[\hat\Pi^{A'A}_i\right]_{a'_i,a_j;a_l,a_s}\, 
  \tilde{\cal A}_{a'_ia_ja_s;a'_ka_la_t}
 \end{eqnarray} 
The elements of the corresponding realigned ${\cal B}$ matrix (see {\ref{abdef}))  are then identified as, 
  \begin{eqnarray} 
 \label{16}
  {\cal B}_{a_u a_s;a_va_t}&=&{\cal A}_{a_u a_v;a_sa_t}\nonumber \\ 
  &=&\sum_{i}\, \sum_{a'_i,a_j,a_l,a_s} \left[\hat\rho^{A}_i\right]_{a_sa_t}\, \left[\hat\Pi^{A'A}_i\right]_{a'_i,a_j;a_l,a_s}\, 
  \tilde{\cal A}_{a'_ia_ja_s;a'_ka_la_t}\nonumber \\ 
\Rightarrow\  {\cal B}&=&\sum_i\  \hat\rho^{A}_i\otimes {\rm Tr}_{A'A}[\hat\Pi^{A'A}_i\,\tilde{\cal A} ]\nonumber \\ 
 &=&\sum_i\ \sum_\alpha\ \hat\rho^{A}_i\otimes {\rm Tr}_{A'A}\left[\hat\Pi^{A'A}_i\ \left( \tau_\alpha^{A'}\otimes P_\alpha^{A}\otimes Q_\alpha^{T}\right) \right]\nonumber \\
&=& \sum_i\ \sum_\alpha\ \, {\rm Tr}_{A'A}\left[\hat\Pi^{A'A}_i\ \left(\tau_\alpha^{A'}\otimes P_\alpha^{A}\right)\right]\, \rho^{A}_i\otimes Q_\alpha^{T}\nonumber \\
{\cal  B}&=&\sum_\alpha \left[\sum_i q_{i\alpha}\, \hat\rho^{A}_i\right]\otimes Q_\alpha^{T}\nonumber \\
&=&\sum_\alpha\ \hat\eta_\alpha^{A}\otimes Q_\alpha^{T}
\end{eqnarray} 
where we have denoted 
\begin{eqnarray}
\label{etadef}
\sum_\alpha q_{i\alpha}\, \hat\rho^{A}_i&=&\hat\eta_\alpha^{A}, \\
\label{qdef}
 q_{i\alpha}={\rm Tr}_{A'A}[\hat\Pi^{A'A}_i\ \left(\tau_\alpha^{A'}\otimes P_\alpha^{A}\right)]. \nonumber 
\end{eqnarray}
 
Now, we consider a specific example of two qubit state (see (\ref{a'ab})) 
\begin{eqnarray}
\label{a'a}
\hat\rho^{A'A}&=&{\rm Tr}_B[\hat\rho_{A'AB}] \nonumber \\ 
&=&  p\, \vert 1_{A'},0_A\rangle\, \langle 1_{A'}, 0_A\vert + (1-p)\,  
\vert 0_{A'}, +_A\rangle\, \langle 0_{A'},+_A,\vert 
\end{eqnarray}
and complete, orthogonal projective measurement (\ref{gmp}).  

We choose the following set $\{P^{A}_\alpha\}$ of  linearly independent $2\times 2$ matrices (see (\ref{12})), which serve as a basis~\cite{RRK} for single qubit systems: 
\begin{eqnarray}
\label{pai}
P^A_1=\frac{1}{2}[I+\sigma_1]=\frac{1}{2}\left(\begin{array}{cc}1 & 1 \\ 1 & 1\end{array}\right), && P^A_2=\frac{1}{2}[I+\sigma_2]=\frac{1}{2}\left(\begin{array}{cc}1 & -i \\ i & 1\end{array}\right) \nonumber \\
P^A_3=\frac{1}{2}[I+\sigma_3]=\left(\begin{array}{cc}1 & 0 \\ 0 & 0\end{array}\right), && P^A_4=\frac{1}{2}[I-\sigma_1]=\frac{1}{2}\left(\begin{array}{cc}1 & -1 \\ -1 & 1\end{array}\right).    
\end{eqnarray}
The corresponding set of Hermitian matrices $\{Q_\beta\}$, which are orthogonal to $\{P^A_\alpha\}$  and obey the property $\sum_\beta Q_\beta=I$ (see Eq. (\ref{13})) are given by, 
\begin{eqnarray}
\label{qai}
Q_1&=&\frac{1}{2}\left(I+\sigma_1+\sigma_2-\sigma_3\right)=\frac{1}{2}\left(\begin{array}{cc}0 & 1-i \\ 1+i & 2 \end{array}\right),  \ Q_2=-\sigma_2=\left(\begin{array}{cc}0 & i \\ -i & 0\end{array}\right) \nonumber \\
Q_3&=&\sigma_3=\left(\begin{array}{cc} 1 & 0 \\ 0 & -1\end{array}\right), \  Q_4=\frac{1}{2}\left(I-\sigma_1+\sigma_2-\sigma_3\right)=\frac{1}{2}\left(\begin{array}{cc} 0 & -1-i \\ -1+i & 2\end{array}\right)    
\end{eqnarray}

Further, choosing  
\begin{equation}
\label{taua'}
\tau^{A'}_{1,4}=\vert 0\rangle_{A'}\langle 0\vert,\ \  \tau^{A'}_{2,3}=\vert 1\rangle_{A'}\langle 1\vert
\end{equation} 
 in (\ref{14}) and simplifying
(using (\ref{a'a}), (\ref{pai}) and (\ref{qai})  we obtain 
\begin{eqnarray}
\label{con} 
\tilde{\cal A}(\rho^{A})=\tilde{\cal A}((1-p)\, P^{A}_1+p\, P^{A}_3)=(1-p)\, P_1 \otimes \tau_1 + 
p\, P_3 \otimes \tau_3=\hat\rho_{A'A}
\end{eqnarray}
confirming the consistency of the assignment map $\tilde{\cal A}$. 

Using the explicit matrices  $\{P^A_\alpha\}$,\  $\{Q_\beta\}$ of (\ref{pai}), (\ref{qai}),  and (\ref{taua'}), along with (\ref{gmp}) for projective measurements,  we  obtain   (see (\ref{qdef}))
\begin{eqnarray}
q_{i1}&=&{\rm Tr}_{A'A}[\hat\Pi^{A'A}_i\ \left( P^{A}_l \otimes \tau_1\right)]=(1,0,0,0)\nonumber \\ 
q_{i2}&=&{\rm Tr}_{A'A}[\hat\Pi^{A'A}_i\ \left(\ P^{A}_2\ \otimes \tau_2\right)]=(0,0,\frac{1}{2},\frac{1}{2})\nonumber \\
q_{i3}&=&{\rm Tr}_{A'A}[\hat\Pi^{A'A}_i\ \left(\ P^{A}_3\ \otimes \tau_3\right)]=(0,0,1,0)\nonumber \\
q_{i4}&=&{\rm Tr}_{A'A}[\hat\Pi^{A'A}_i\ \left(\ P^{A}_4\ \otimes \tau_4\right)]=(0,1,0,0)
\end{eqnarray}
and (see ((\ref{rhoai}), \ref{etadef}))
\begin{eqnarray}
\label{30}
\eta_{1}&=&\sum_i\, q_{i1}\,\rho_i^{A}=\sum_i\, q_{i1}\, {\rm Tr}_{E}[\hat\Pi^{A'A}_i]={\rm Tr}_{E}[\hat\Pi^{A'A}_1]=\vert +\rangle\langle +\vert=P_1\nonumber \\
\eta_{2}&=&\sum_i\, q_{i2}\,\rho_i^{A}=\sum_i\, q_{i2}\, {\rm Tr}_{E}[\hat\Pi^{A'A}_i]=\frac{1}{2}\left({\rm Tr}_{E}[\hat\Pi^{A'A}_3]+{\rm Tr}_{E}[\hat\Pi^{A'A}_4]\right)\nonumber \\ 
&=&\frac{1}{2}(\vert 0\rangle\langle 0\vert+\vert 1\rangle\langle 1\vert)=\frac{I}{2} \nonumber\\
\eta_{3}&=&\sum_i\, q_{i3}\,\rho_i^{A}=\sum_i\, q_{i3}\, {\rm Tr}_{E}[\hat\Pi^{A'A}_i]={\rm Tr}_{E}[\hat\Pi^{A'A}_3]=\vert 0\rangle\langle 0\vert=P_3\nonumber \\
\eta_{4}&=&\sum_i\, q_{i4}\,\rho_i^{A}=\sum_i\, q_{i4}\, {\rm Tr}_{E}[\hat\Pi^{A'A}_i]={\rm Tr}_{E}[\hat\Pi^{A'A}_2]=\vert -\rangle\langle -\vert=P_4
\end{eqnarray}

We thus obtain the ${\cal B}$ map (see (\ref{16})) corresponding to this particular example as, 
\begin{eqnarray}
\label{31}
{\cal B}&=&P_1\otimes Q_1 + \frac{I}{2}\otimes Q_2+P_3\otimes Q_3 + P_4\otimes Q_4\nonumber \\ 
&=& \left(\begin{array}{cccc} 1 & 0 & 0 & \frac{1}{2} \\ 
0 & 0 & \frac{1}{2} & 0 \\
0 & \frac{1}{2} & 0 & 0  \\
\frac{1}{2} & 0 & 0 & 1 
\end{array}\right).
\end{eqnarray}


\vspace*{-5pt}   

\end{document}